\documentclass[a4paper, 12pt]{article} 
\usepackage[right=1in,left=1in,top=1in,bottom=1in]{geometry}
\usepackage{amsfonts,amsmath,amssymb,amsthm,mathrsfs,url,hyperref, graphicx, pxfonts}
\usepackage{setspace}  
\usepackage[usenames,dvipsnames]{xcolor}
\usepackage[T1]{fontenc} 
\usepackage{natbib}
\usepackage{epstopdf}
\usepackage{enumitem}
\usepackage{sectsty}
\usepackage{wrapfig} 
\sectionfont{\large}
\bibpunct{(}{)}{,}{a}{}{,}
\usepackage{tcolorbox}
    \usepackage{pgfplots}

\usepackage{tikz}   
\usepackage{tikz-3dplot} 

\usepackage{bbm}
\usepackage{MnSymbol}
\usepackage[all]{xy}

\usepackage{epigraph}
\usepackage[normalem]{ulem} 

\newcommand{\PH}{0} 
\newcommand{\macro}{0} 
\newcommand{\Hilbert}{\mathscr{H}}

\newcommand{\sE}{\mathscr{E}}

\newcommand{\sS}{\mathscr{S}}

\newcommand{\sZ}{\mathscr{Z}}

\newcommand{\EEE}{\mathbb{E}}

\newcommand{\PPP}{\mathbb{P}}
\newcommand{\RRR}{\mathbb{R}}

\newcommand{\SSS}{\mathbb{S}}

\newcommand{\ket}[1]{|#1\rangle}
\newcommand{\bra}[1]{\langle #1|}
\DeclareMathOperator{\tr}{tr}
\newcommand{\sphere}{\mathbb{S}}

\newtheorem{thm}{Theorem}
\newtheorem{cor}{Corollary}
\theoremstyle{definition}
\newtheorem{dfn}{Definition}
\newtheorem{ex}{Example}

\newcommand{\EC}[1]{{\color{black}#1}}
\newcommand{\RT}[1]{{\color{black}#1}}

\newcommand{\be}{\begin{equation}}
\newcommand{\ee}{\end{equation}}

\makeatletter

\newlength{\bibitemsep}
\newlength{\bibparskip}
\let\oldthebibliography\thebibliography
\renewcommand\thebibliography[1]{%
  \oldthebibliography{#1}%
  \setlength{\parskip}{\bibitemsep}%
  \setlength{\itemsep}{\bibparskip}%
}

\title{\textbf{Typical Quantum States of the Universe are \\ Observationally Indistinguishable}}

\author{
Eddy Keming Chen\footnote{Department of Philosophy, University of California San Diego, 9500 Gilman Dr, La Jolla, CA 92093, USA.  Email: eddykemingchen@ucsd.edu}~~and
Roderich Tumulka\footnote{Fachbereich Mathematik,
     Eberhard-Karls-Universit\"at T\"ubingen, Auf der Morgenstelle 10, 72076
     T\"ubingen, Germany. Email: roderich.tumulka@uni-tuebingen.de}
}

\date{\emph{British Journal for the Philosophy of Science}, Accepted Version}

\begin{document}
\bibliographystyle{apa}

\maketitle 

Abstract: We establish three impossibility results regarding our knowledge of the quantum state of the universe. Suppose the universal quantum state is a typical unit vector in a high-dimensional subspace $\Hilbert_0$ of Hilbert space $\Hilbert$, such as the low-entropy subspace defined by the Past Hypothesis. We show that: (1)  Any particular observation is incapable of identifying the universal state vector in $\Hilbert_0$ or substantially reducing the set of possibilities. In other words, the overwhelming majority of possible state vectors are observationally indistinguishable from each other. (2) For any reasonably probable measurement outcome and for most pairs of vectors in $\Hilbert_0$, that outcome will not appreciably favor one vector over the other. (3) Bayesian updating on any measurement result, unless it is extraordinarily improbable, has a negligible effect on the initial uniform probability distribution over the states in $\Hilbert_0$. These findings represent the most stringent epistemic constraints known for a quantum universe and are derived from a typicality theorem in quantum statistical mechanics. We close by considering how theoretical considerations beyond empirical evidence might inform our understanding of this fact and our knowledge of the universal quantum state.

\bigskip

Key words: limitation to knowledge; empirical equivalence; past hypothesis; quantum statistical mechanics; quantum measurement; measure concentration.


\begingroup
\singlespacing
\tableofcontents
\endgroup


\section{Introduction}

We establish three impossibility results regarding our knowledge of the quantum state of the universe---a central object in \EC{several leading formulations of} quantum theory. We show that typical universal quantum states are observationally indistinguishable, in the sense that no observation can distinguish typical universal quantum states in a high-dimensional subspace of the Hilbert space, such as the one prescribed by the Past Hypothesis (see \S\ref{sec:PH} and \citep{Alb00,goldstein2019gibbs}), unless we somehow already knew the quantum state of the universe. We call this fact \textit{observation typicality}.

Observation typicality follows from a result in quantum statistical mechanics called \textit{distribution typicality} \citep{reimann2007typicality,teufel2023time}: for any observable, most quantum states in a high-dimensional Hilbert subspace $\Hilbert_0$ lead to very nearly identical probability distributions for that observable. Roughly speaking, that is, most wave functions from a high-dimensional $\Hilbert_0$ look the same.

We assume here that there is a wave function $\Psi_t$ of the universe and that at the initial time $t_0$ of the universe (say, at the big bang), $\Psi_{t_0}$ had to lie in a particular subspace $\Hilbert_{\macro}$ of the Hilbert space $\Hilbert$ of the universe (say, corresponding to a low-entropy macro-state, \S\ref{sec:PH}). Observation typicality then asserts that our empirical data at any time $t$ will reveal very little about $\Psi_{t_0}$ (beyond its membership in $\Hilbert_{\macro}$, which we take as known), even if we assume that any observable (including any POVM) can be measured.

Observation typicality is a surprisingly strong result, yet it has gone unnoticed so far. There are other known examples of epistemic limitations in physics. \cite{durr2004quantum} already proved that it is impossible to measure a wave function precisely. Observation typicality takes this limitation a step further, delivering an even stronger conclusion: typically, even an imprecise measurement is impossible. No observation---provided it does not disconfirm the physical theory---can yield substantive information about the quantum state. In this sense, the limitation to knowledge is more pervasive and persistent in a quantum universe than previously recognized.

We note that a statement about most quantum states in the \emph{full} Hilbert space $\Hilbert$ of the universe would have limited applicability because it may fail for the quantum states of interest. For example, most quantum states in $\Hilbert$ are in thermal equilibrium, but the actual quantum state of our universe is not. To use typicality theorems in our situation, we apply them to a \emph{subspace} $\Hilbert_{\macro}$ of $\Hilbert$ such as the subspace comprising the quantum states $\Psi$ compatible with a certain macro-state of the universe that may be far from thermal equilibrium. We assume that $\Hilbert_{\macro}$ has finite but huge dimension $d_{\macro}$ (while $\Hilbert$ may have finite or infinite dimension).\footnote{If our universe initially had a finite volume and occupied a finite energy interval $[E_1,E_2]$, it is confined to a Hilbert space $\Hilbert_{[E_1,E_2]}$ of finite (albeit realistically very large) dimension. See \S\ref{sec:PH}.} We can use the result while accepting certain physical laws, including those pertaining to initial conditions of the universe, according to which our universe is a typical member of the possible universes compatible with such laws.

It is well known that if we pick two vectors $\Psi_A,\Psi_B$ purely randomly from the unit sphere of a high-dimensional $\Hilbert_0$, then with overwhelming probability they will be nearly orthogonal, and thus there exists a quantum observable $P$ that distinguishes them reliably \RT{\citep{Helstrom1976}}. However, this fact does not affect the question we are studying here because the choice of $P$ depends on (at least one of) $\Psi_A,\Psi_B$, while we do not assume we already know the wave function of the universe. The situation that we do not allow the observable to depend on the quantum state is reflected in formal statements by the order of
quantifiers such as \textit{some}, \textit{most}, and \textit{every}. 
To clarify the order of quantifiers in our result: By observation typicality, we mean that for every observation, most quantum states are not distinguishable by that observation. 
The alternative order---where we first quantify over \textit{most} states and then over \textit{every} observation---would imply an even stronger claim, which we call \textit{super-strong observation typicality}. The latter claim is generally false, though it holds for \RT{the wave function of the universe if (as is the case for humans) only local observations in a small part of the universe are possible---}because of the more well-known result of canonical typicality, as discussed in \cite{chentumulka2025b}.

In classical mechanics, the statement analogous to observation typicality fails. It holds in quantum mechanics because, roughly speaking, quantum states can be in superpositions, and the superposition weights of any particular observation given by typical quantum states are nearly identical. The latter is a version of the mathematical fact known as the concentration-of-measure phenomenon (the situation where random quantities become nearly deterministic) and occurs in high-dimensional Hilbert spaces. We note further that the same arguments used to establish observation typicality yield that, if the time evolution is unitary, also $\Psi_t$ (instead of $\Psi_{t_0}$) cannot be distinguished empirically from typical alternative vectors in the appropriate subspace, i.e., in $U_t\Hilbert_{\macro}$ with $U_t= \exp(-iH(t-t_0))$ and $H$ the Hamiltonian of the universe.

Our conclusions are based on three mathematical results.  First (\S\ref{sec:reliably}), that a typical unit vector $\Psi_{t_0}\in\Hilbert_{\macro}$ cannot be reliably distinguished from the density matrix $\rho_{\macro}$ associated with a uniform distribution in $\Hilbert_{\macro}$, i.e., $\rho_{\macro}=P_{\macro}/d_{\macro}$ with $P_{\macro}$ the projection operator to $\Hilbert_{\macro}$ and $d_{\macro}=\dim \Hilbert_{\macro}$. Consequently, typical unit vectors cannot be reliably distinguished from each other. By contrast, the second result (\S\ref{sec:unreliably}) is in a sense stronger but does require that the observation in question be not too improbable. It asserts that \emph{even unreliably}, the states cannot be distinguished, which conveys that such empirical data do not even yield (any significant amount of) partial or probabilistic information about $\Psi_{t_0}$. 
The third result (\S\ref{sec:Bayesian}) is expressed in Bayesian terms: if we start from the uniform  prior $u_{\macro}$ over the unit sphere $\SSS(\Hilbert_{\macro})$ (containing all normalized wave functions) in $\Hilbert_{\macro}$ 
and update in a Bayesian way upon an observation that is not too improbable, then the posterior distribution will still be very nearly uniform over $\SSS(\Hilbert_{\macro})$.

Some empirical observations might, of course, change our mind about which subspace $\Hilbert_\PH$ is the correct one. Since we regard here the specification of $\Hilbert_\PH$ (in the Past Hypothesis) as part of the given theory, such an observation would amount to a disconfirmation of the theory. This leads to the interesting question of which kind of observations should be regarded as disconfirming the theory (see \citep{chentumulka2025b}).  
We take here for granted that the theory is correct and not getting disconfirmed, and remark only briefly that if the specification of $\Hilbert_\PH$ has the status of a law of nature, we may expect it to be simple and natural in order to be convincing \citep{loewer2024OUP, chen2024CUP}; as a consequence, we would not be inclined to adjust the choice of $\Hilbert_\PH$ in complicated ways just to adapt it to some random-looking features of our world (such as, e.g., the shape of Ireland). 

\RT{There are various other reasons limiting our epistemic access to the universal quantum state; for example, since the wave function of the universe undergoes ``branching'' at every measurement and other instances of decoherence, we should expect we can only access the branch we are in. Our result, however, provides a particularly strong limitation in a clean and rather simple way from just one reason: the high dimension of $\Hilbert_0$.} \EC{We return to this issue in \S3.1.}

Our results show that insofar as we have detailed knowledge about the universal quantum state, it is even more theoretical than has been recognized.  They should be of interest to anyone interested in the epistemological implications of physics. 

\RT{Observation typicality applies not only to the universe as a whole but in fact also to any quantum system with a Hilbert space $\Hilbert$ of high dimension if the wave function $\Psi_{t_0}$ at some time $t_0$ is typical in a subspace $\Hilbert_0$ of high dimension $d_0$.\footnote{\RT{The situation can be different if we are provided with $n\gg 1$ copies of this system and know that all copies have the \emph{same} wave function $\Psi_{t_0}$: if $n\gg d_0^2$, we could determine $\Psi_{t_0}$ approximately with high probability by measuring different observables on different copies.}} This aspect is explored further in \cite{chentumulka2025b}.}

This paper is organized as follows. In \S\ref{sec:distrtyp}, we formulate distribution typicality and explain how it applies to the special case of the Past Hypothesis. In \S\ref{sec:oi}, we clarify our criterion for observational indistinguishability, derive observation typicality from distribution typicality, and provide a Bayesian analysis. In \S\ref{sec:known}, we compare observation typicality to known results about epistemic limitations in physics. In \S\ref{sec:implications}, we discuss potential philosophical implications. 
In \S\ref{sec:conclusion}, we conclude.

\section{Distribution Typicality}
\label{sec:distrtyp}

In the 21st century, many exciting new results have been proven in quantum statistical mechanics and quantum thermodynamics, leading to improved understanding of the thermodynamic behavior of a closed macroscopic quantum system in a pure state \citep{gemmer2009quantum,tas16,Mori18}. Such results often assume an ``individualist'' attitude, according to which an individual quantum system in  a pure state, as opposed to either a density matrix or an ensemble of systems, is studied with respect to its thermodynamic properties. These results often concern \textit{typical properties} of those states.\footnote{The concept of typicality and its role in physical theories have received growing attention in the philosophical literature. For some recent examples, see \cite{goldstein2012typicality}, \cite{LazaroviciReichert2015}, \cite{Allori2020}, \cite{Wilhelm2022}, \cite{Lazarovici2023}, \cite{Hubert2024}, \cite[sec.\ 4.5]{FriggSEP2024}, and the references therein. \RT{We use the convenient language of probability theory even when we talk about events that cannot be repeated such as the choice of the initial wave function $\Psi$ of the universe. Despite this language, our results are equally valid when the distributions for $\Psi$ (or local beables $X$) of the whole universe are understood in the sense of \emph{typicality}, essentially because there $\Psi$ (or $X$) still behaves \emph{as if} randomly chosen.
} }

We shall focus on \textit{distribution typicality} in this paper.  In \S\ref{sec:oi}, we show that distribution typicality implies observation typicality for orthodox quantum mechanics (OQM), Bohmian mechanics (BM), the Ghirardi-Rimini-Weber spontaneous collapse theory (GRW), and Everettian quantum mechanics (EQM) \citep{Barrett2019,DT09,Ghi07,BarrettSEP2023}. In this section, we formulate distribution typicality and specialize it to the Past Hypothesis. 

\subsection{The Key Theorem}
\label{sec:keythm}

\textbf{Terminology.}
Recall that $\SSS(\Hilbert_{\macro})$ is the unit sphere in the subspace $\Hilbert_{\macro}$, $d_{\macro} = \dim \Hilbert_{\macro}$, and $u_{\macro}$ the uniform probability distribution over $\SSS(\Hilbert_{\macro})$; that is, for any subset $S$ of $\SSS(\Hilbert_{\macro})$, $u_{\macro}(S)$ is the surface area of $S$, normalized by dividing through the surface area of $\SSS(\Hilbert_{\macro})$. We say that some statement about the wave function $\Psi$, $s(\Psi)$, is true \emph{for $(1-\varepsilon)$-most $\Psi\in\sphere(\Hilbert_{\macro})$} if and only if the set $S$ of $\Psi\in\sphere(\Hilbert_{\macro})$ satisfying $s(\Psi)$ has $u_{\macro}(S)\geq 1-\varepsilon$ (equivalently, if $S$ has at least the fraction $1-\varepsilon$ of the surface area, or if a purely random point on the sphere has probability $\geq 1-\varepsilon$ to lie in $S$). Averages on $\sphere(\Hilbert_{\macro})$ will also be taken with respect to $u_{\macro}$.\footnote{The results in this paper can be extended to  the Gaussian adjusted projected (GAP) measures introduced by \cite{goldstein2006distribution}, as is done by \cite{teufel2024canonical}. But here, we consider only the uniform measure $u_{\macro}$.} We also need the concept of POVM.

\textbf{POVM.} A POVM (positive-operator-valued measure) is a generalization of the notion of an observable represented by a self-adjoint operator. In the simple case with \emph{discrete} outcomes, a POVM on a Hilbert space $\Hilbert$ is a family of positive operators $E_z$ on $\Hilbert$ that add up to the identity operator. More generally, a POVM is a map $E$ from the measurable subsets of an outcome space $\sZ$ to the set of positive operators on $\Hilbert$ satisfying normalization, $E(\sZ)=I$, and countable additivity. POVMs have an important representational role in quantum mechanics, given by the Main Theorem about POVMs: \textit{for every quantum physical experiment $\sE$ on a quantum system $\sS$ whose possible outcomes lie in a space $\sZ$, there exists a POVM E on $\sZ$ such that, whenever $\sS$ has wave function $\Psi$ at the beginning of $\sE$, the random outcome $Z$ has probability distribution given by}
\be
\PPP (Z=z) = \bra{\Psi}\,E_z\,\ket{\Psi}.
\ee
\textit{where $E_z$, the image of $\{z\}$ under $E$, is an element of the POVM $E$.} See \cite{durr2004quantum} and \cite{tumulka2022foundations} for proofs. 
 
The Main Theorem about POVMs provides a generalization of the Born rule to arbitrary experiments (instead of ideal quantum measurements). The operator $E_z$ is associated with the possible outcome $z$; in the case of an ideal quantum measurement, $E_z$ would be the projection operator to the eigenspace of the eigenvalue $z$.

We can now state the key theorem as follows: 

\begin{thm}[Distribution Typicality]\label{thm:1}
Let $\Hilbert_{\macro}$ be an arbitrary subspace with finite dimension $d_{\macro}$ in $\Hilbert$, let the density matrix $\rho_{\macro}=P_{\macro}/d_{\macro}$ be the normalized projection to $\Hilbert_{\macro}$, and let $B$ be any self-adjoint operator on $\Hilbert$. Then for every $\varepsilon>0$, for $(1-\varepsilon)$-most $\Psi\in\sphere(\Hilbert_{\macro})$,
\be\label{prop1Abound1}
\Bigl|\bra{\Psi}\,B\,\ket{\Psi} - \tr(\rho_{\macro} \, B)\Bigr| \leq \frac{\|B\|}{\sqrt{\varepsilon d_{\macro}}} \,,
\ee  
where the operator norm $\|B\|$ means the largest absolute eigenvalue of $B$. Moreover, for every POVM $\{E_1,\ldots,E_L\}$ and every $\varepsilon>0$, for $(1-\varepsilon)$-most $\Psi\in\sphere(\Hilbert_{\macro})$, the distribution $\langle\Psi|E_z|\Psi\rangle$ is close to the distribution $\tr(\rho_{\macro} \, E_z)$ in the sense that for each $z$,
\be\label{prop1Abound}
\Bigl|\bra{\Psi}\,E_z\,\ket{\Psi} - \tr(\rho_{\macro} \, E_z)\Bigr| \leq \sqrt{\frac{L}{\varepsilon d_{\macro}}} \,.
\ee  
\end{thm}

Note that the right-hand side of \eqref{prop1Abound} is small whenever
\be\label{dmuepsilon}
d_{\macro} \gg \frac{L}{\varepsilon} \,.
\ee
Theorem \ref{thm:1} is a particular case of Theorem 3 proven by \cite{teufel2023time}. It can also be derived from a theorem proven by \cite{reimann2007typicality}. \RT{It is an application of Chebyshev's inequality to the random variable $X:=\langle \Psi|B|\Psi\rangle$ with $\Psi$ uniformly distributed over the unit sphere, together with bounds on the variance of $X$ \cite[Sec.~4]{teufel2023time}. (A tighter bound could be obtained from L\'evy's lemma \cite[Sec.s 5.1 and 8.1]{teufel2023time}, but the difference will not change the further reasoning in a relevant way.)} 
\EC{Inequality \eqref{prop1Abound} follows from inequality \eqref{prop1Abound1}.}\footnote{Here is how: Replace $B$ by $E_z$ and $\varepsilon$ by $\varepsilon/L$ in the statement involving \eqref{prop1Abound1} and use $\|E_z\|\leq 1$. Thus, for any $z$, the set of $\Psi$'s satisfying \eqref{prop1Abound} has measure at least $1-\varepsilon/L$, so the intersection of these sets has measure at least $1-\varepsilon$.}

It can be shown that $\tr(\rho_{\macro} E_z)$ is exactly the average of $\langle\Psi|E_z|\Psi\rangle$ over $\SSS(\Hilbert_{\macro})$ (using the uniform distribution $u_{\macro}$); thus, \eqref{prop1Abound} expresses that $\langle\Psi|E_z|\Psi\rangle$ is close to its average value
whenever \eqref{dmuepsilon} holds.
Theorem~\ref{thm:1} is an example of the ``concentration-of-measure phenomenon.'' 
 Here, we have that for any observation represented by a POVM, typical quantum states in a high-dimensional Hilbert space have very nearly identical Born probability for that observation. Let us clarify and elaborate on this theorem. 

\begin{ex}
\label{ex:1}
Which size should we consider for $d_0$ and $\varepsilon$ in practice? 
For a macroscopic quantum system with $N$ particles, the dimension of an energy shell is usually of order $10^N$ (or perhaps more like $10^{1000 \,N}$). While the dimensions of subspaces corresponding to different macro states differ by huge factors, even the comparatively small subspaces of this kind, corresponding to states far from thermal equilibrium, have dimension of order roughly $10^N$. For the universe, a widespread estimate is $N=10^{80}$, which motivates us to consider,
\be\label{exd0}
\text{for example, }d_0= 10^{10^{80}}\,.
\ee
Let us turn to $\varepsilon$. \cite{Bor62} famously argued that an event with probability,
\be\label{exepsilon1}
\text{for example, }\varepsilon=10^{-200}
\ee
or smaller can be assumed to never occur in the history of the universe. Of course, we can make sure that some event of smaller probability occurs, for example by tossing a fair coin 1000 times; then the actual sequence of heads and tails had probability $2^{-1000}\approx 10^{-300}$ like every sequence; but Borel meant that any event \emph{specified in advance} with probability $\leq 10^{-200}$ is not going to occur. Correspondingly, for this value of $\varepsilon$ and any given POVM, we can expect $\Psi_{t_0}$ to lie in the set of $\Psi$'s satisfying \eqref{prop1Abound}.
While our results do not require the values \eqref{exd0}, \eqref{exepsilon1}, we will use them for illustration. So for our purposes, \eqref{dmuepsilon} is satisfied, for example for $L=2$ or $L=1000$, with enormous margins. Our results in \S\ref{sec:oi} will be based on \eqref{prop1Abound1}, as it provides the best bound for a given outcome $z$.
\end{ex}

\begin{ex}
\label{ex:2}
The high dimension is relevant for the typicality statement. In a 2-dimensional Hilbert space $\Hilbert_0=\Hilbert=\mathbb{C}^2$ such as for a spin-1/2 particle with fixed position, 10\%\ of all unit vectors (a substantial fraction) have 90\%\ or greater probability (far from the average 50\%) for the outcome ``up'' in a $z$-spin measurement,\footnote{\label{fn:ex2}More generally, for any $p\in[0,1]$, the fraction $p$ of all $\psi$'s in $\mathbb{C}^2$ have probability $\geq 1-p$ for ``up.'' Proof: With every unit spinor $\psi$ there is associated a real unit 3-vector with components $n_i=\langle\psi|\sigma_i|\psi\rangle$, and the uniform distribution over unit spinors leads to a uniform distribution over real unit 3-vectors. For the outcome $Z$ of a $z$-spin measurement, $\PPP(Z=\mathrm{up})=\langle\psi|P_{\uparrow}|\psi\rangle$, where $P_{\uparrow}=|\uparrow\rangle\langle\uparrow|=\frac{1}{2}(I+\sigma_z)$ is the projection to the $+1$-eigenspace of $\sigma_z$, so $\PPP(Z=\mathrm{up})= \frac{1}{2}+\frac{1}{2}n_z$. In spherical coordinates, $n_z=\cos \theta_0$, and the set $M\subset\SSS(\mathbb{R}^3)$ of $\boldsymbol{n}$'s with $\theta\leq \theta_0$ has area $\int_0^{2\pi}d\varphi \int_0^{\theta_0} d\theta\, \sin \theta=2\pi(1-\cos \theta_0)$, so the fraction of $\psi$'s with $\boldsymbol{n}\in M$ is area$(M)/4\pi= \tfrac{1}{2}(1-\cos\theta_0)$ = $1-\PPP(Z=\mathrm{up})$.} whereas in the Hilbert space $\Hilbert=(\mathbb{C}^2)^{\otimes N}$ of $N$ spin-$1/2$ particles with fixed positions and large $N$, nearly 100\%\ of the $\psi$'s have probability near 50\%\ for the outcome ``up'' in a $z$-spin measurement on particle 1. If we took $\psi$ to factorize, $\psi=\psi_1 \otimes \psi_2 \otimes \cdots \otimes \psi_N$, with each $\psi_i$ uniformly distributed, then the probability of outcome ``up'' for particle 1 would still be 90\%\ or greater in 10\%\ of the cases; but most $\psi$ do not factorize but are highly entangled. 
\end{ex}

\textbf{Minimal Assumptions.} Theorem \ref{thm:1}  holds in great generality. First, it applies to an arbitrary POVM. (In fact, it even applies to every bounded operator.) By the Main Theorem about POVMs, the theorem covers every  possible outcome arising from an arbitrary quantum physical experiment. Second, it does not make assumptions about the interaction between the subsystem and the environment. Third, it does not invoke chaos, ergodicity, or mixing. The typicality in distribution typicality comes from the typicality of quantum states in $\sphere(\Hilbert_{\macro})$ for a given observation, and the large number comes from $d_{\macro}$, the large dimension of $\Hilbert_{\macro}$.  Finally, distribution typicality applies to thermal equilibrium and non-equilibrium. For concreteness, in \S\ref{sec:PH} we provide a physical interpretation for a universe in thermal non-equilibrium, taking $\Hilbert_{\macro}$ to be the low-entropy initial macro-state of the universe prescribed by the Past Hypothesis.  

\textbf{Difference from the Classical Case.} We do not have distribution typicality in classical mechanics. The prediction of any observation from a particular \textit{micro-state} in the classical phase space has a trivial probability of $0$ or $1$, while for many observables their probability distribution induced by the uniform probability distribution in a region of phase space (the ensemble statistics) is non-trivial. 
For example, we can design a coin-flip experiment where half the microstates assign probability $0$ and the other half assign probability $1$ to the ``Tails'' outcome. 
 The relevant way in which quantum mechanics differs from classical mechanics is that if a wave function $\Psi=\Psi_{t_0}$ gets chosen randomly, then the outcome $Z$ of an observation is ``doubly random'': Given $\Psi$, it is random with the Born distribution $\langle \Psi|E_z|\Psi\rangle$, and in addition $\Psi$ is itself random. The crucial point is that in our situation, different $\Psi$s have very similar superposition weights, with the consequence that they have nearly identical $\langle \Psi|E_z|\Psi\rangle$.

\subsection{Main Motivation: The Past Hypothesis}
\label{sec:PH}

A typicality statement about most quantum states in some space has limited applicability if the relevant quantum states are atypical in that space. For example, most quantum states in the energy shell are in thermal equilibrium, but many quantum states we observe are not. That is not a problem in our case, as Theorem \ref{thm:1} is quite general. To apply Theorem \ref{thm:1} to the relevant class of quantum states, we can specialize the arbitrary subspace $\Hilbert_{\macro}$ to any macro-state that may be far from  thermal equilibrium. We are particularly motivated by the thought that the physical laws may require the initial wave function of the universe to lie in a particular subspace $\Hilbert_\PH$.

In fact, it has been suggested for the explanation of the thermodynamic arrow of time \citep[p.~115]{Fey63}, \citep{Pen79,Leb93,Alb00,goldstein2019gibbs, chen2018IPH} that the initial state of the universe must be restricted to a set of low-entropy states; in a quantum theory, such a set would be given by a suitable subspace $\Hilbert_0$ of the Hilbert space $\Hilbert$ of the universe, and the presumed additional law might then be formulated as follows:

\begin{description}
    \item[Past Hypothesis (PH)] $\Psi_{t_0}$ is a typical element of $\mathscr{H}_{\PH}$.
\end{description}

Here, $\Hilbert_\PH$ is assumed to contain all wave functions compatible with a certain macro-state. For example, Penrose's Weyl curvature hypothesis might amount to taking as $\Hilbert_\PH$ something like the joint eigenspace with all eigenvalues 0 of all Weyl curvature operators at $t_0$. To explain the arrow of time, the initial macro-state should have very low entropy. Taking as the definition of entropy the  quantum Boltzmann entropy \citep{Leb93,GLTZ10,goldstein2019gibbs}
\be
S_{qB} (\Psi_{t_0}) = k_B \text{log} (d_{\PH} ) \,,
\ee
the condition of low entropy corresponds the condition that the dimension $d_\PH$ of $\Hilbert_\PH$ is much smaller than the dimension of the full Hilbert space $\Hilbert$, in fact much smaller than the dimension of the subspace $\Hilbert_{eq}$ corresponding to thermal equilibrium in the same energy shell, $d_{\PH} \ll d_{eq}$. Then the PH requires $\Psi_{t_0}$ to have low entropy and to be far from thermal equilibrium.

With respect to the normalized uniform measure $u_{\PH}$, we expect that typical wave functions in $\sphere(\mathscr{H}_{\PH})$ will evolve in a way such that  $S_{qB}$ increases in the medium and the long run, satisfying the Developmental Conjecture formulated by \cite{goldstein2019gibbs}, a version of the Second Law of Thermodynamics for a quantum universe. 
In the philosophical literature, $\rho_{\PH}$ has been called the \textit{Wentaculus} density matrix, corresponding to the initial state of the Wentaculus theory \citep{chen2018IPH, chen2024wentaculus}.

\section{Observation Typicality}
\label{sec:oi}

In this section, we show that distribution typicality implies observation typicality. 

\subsection{General Distinctions}

There are various kinds of observational indistinguishability (OI). \EC{While in-practice OI arises from resource limitations, in-principle OI represents fundamental limits that no technological advance could overcome. Though discussions in foundations of physics often focus on exact equivalence of predictions---such as equivalent probability distributions in quantum theory---approximate equivalence can also yield in-principle OI. We focus on proving in-principle limits, which naturally imply in-practice ones.}

\EC{Observation typicality is an in-principle limitation: our results apply to any POVM on the universal Hilbert space, whether implementable with current technology or fantastical capabilities that could measure across branches or the entire universe. While spatial locality and decoherence already limit which observations we can perform, these alone do not entail observation typicality---different universal states could assign vastly different probabilities to events on our branch or region. What we show is that no substantial information about $\Psi_{t_0}$ can be obtained from \emph{any} experiment using \emph{any} resources, whenever the dimension $d_0$ of $\Hilbert_0$ is sufficiently large.} It may seem surprising that an \emph{in-principle} limitation can follow from an \emph{approximate} equality (such as $\langle\Psi|E_z|\Psi\rangle\approx \tr(\rho_0 E_z)$), for two reasons. First, it would seem that nearby quantities can be distinguished better by measuring them with higher resolution; but here, the quantities are, not measurement outcomes, but probabilities of outcomes. Second, it would seem that probabilities can be measured more accurately if we repeat the experiment sufficiently often; but here, that is not possible, as we will discuss in \S\ref{sec:remarks}.

\subsection{Mathematical Descriptions}
\label{sec:math}

In \S\ref{sec:phys}, we will discuss for OQM, BM, GRW, and EQM which limitations for observers arise and how. But first, in this section, we describe the mathematical facts on which theses conclusions rest.

\subsubsection{Not Reliably Distinguishable}
\label{sec:reliably}

We first show that, as a consequence of Theorem~\ref{thm:1}, typical vectors in $\Hilbert_0$ are not \emph{reliably} distinguishable.

Let us make precise what we mean by ``reliable.'' Suppose we use outcome  $z$ to determine whether $A$ is the case or $B$ is the case. We say that $E_z$ ``$(1-\delta)$-reliably'' distinguishes $A$ from $B$ just in case $|\PPP_A(Z=z)-\PPP_B(Z=z)|\geq 1-\delta$, where $\PPP_A$ means the probability if $A$ is the case, and likewise for $\PPP_B$. Thus, a reliability of 95\% requires that the two probabilities differ by at least $95\%$. The reliability lies between 0 and 1, where 1 means perfectly reliable and 0 means completely unreliable.\footnote{Under the decision rule “if $z$ occurs, guess $A$; otherwise guess $B$,” and assume $\PPP_A(z)\ge \PPP_B(z)$.  Then $|\PPP_A(z)-\PPP_B(z)|\geq 1-\delta$ 
implies $\text{Sensitivity} = \PPP_A(z) \geq 1-\delta,$ and $\text{Specificity} = 1 - \PPP_B(z) \geq 1-\delta,$ so the Type I error (false‐positive) rate $\PPP_B(z) \leq \delta$ and the Type II error (false‐negative) rate $1 - \PPP_A(z) \leq  \delta$. Hence, smaller $\delta$ means smaller Type I and Type II errors in hypothesis testing.} 

\begin{dfn}
Let $\{E_z:z\in\sZ\}$ be any POVM, $z$ any element of $\sZ$, and $\delta>0$. We say $E_z$ \emph{$(1-\delta)$-reliably distinguishes the density matrices $\rho_A,\rho_B$} if and only if $|\tr(\rho_A E_z) - \tr(\rho_B E_z)| \geq 1-\delta$.
\end{dfn}

Note that distinguishing at a reliability level $1-\delta$ also counts as distinguishing at any lower reliability level (i.e., larger $\delta$). \RT{(We remark in passing that Helstrom's (\citeyear{Helstrom1976}) theorem provides the optimal POVM for given $\rho_A,\rho_B$ maximizing the reliability level.)}

\begin{ex}
Consider again Example~\ref{ex:2}, with $E_z = P_{\uparrow}=\frac{1}{2}(I+\sigma_z)$. If $\delta$ is taken to be $0.1, 0.2, 0.3$, then $E_z$ $(1-\delta)$-reliably distinguishes, respectively, $1\%, 4\%, 9\%$ of the total pairs of $\psi$'s in $\mathbb{C}^2$. (These fractions are given by $\delta^2$.\footnote{Proof: Let $p_A=\langle\Psi_A|P_\uparrow|\Psi_A\rangle$ and likewise $p_B$. The set of $(p_A,p_B)\in [0,1]^2$ with $|p_A-p_B|\geq 1-\delta$ is the union of the triangles $(0,1-\delta)-(\delta,1)-(0,1)$ and $(1-\delta,0)-(1,\delta)-(1,0)$, which has area $\delta^2$. By Footnote~\ref{fn:ex2}, the set of $(\Psi_A,\Psi_B)$ with such $(p_A,p_B$) has the same measure.}) 
\end{ex}

We can show that for fixed $\delta$, in higher dimensional Hilbert spaces, Theorem~\ref{thm:1} forces the fraction of pairs that $E_z$ $(1-\delta)$-reliably distinguishes to drop quickly and become essentially zero. 

\begin{cor}\label{cor:reliably}
Let $\{E_z:z\in\sZ\}$ be any POVM, $z$ any element of $\sZ$, and $\varepsilon>0$. If $d_0 >  4 /\varepsilon$, then for $(1-\varepsilon)$-most $\Psi_A\in\SSS(\Hilbert_0)$, $E_z$ fails to $\frac{1}{2}$-reliably distinguish $\rho_A := |\Psi_A\rangle\langle\Psi_A|$ from $\rho_B := \rho_0$. Furthermore, If $d_0 > 16/\varepsilon$, then for $(1-\varepsilon)^2$-most choices of $\Psi_A,\Psi_B$, $E_z$ fails to $\frac{1}{2}$-reliably distinguish $\rho_A$ from $\rho_B := |\Psi_B\rangle\langle\Psi_B|$.
\end{cor}

This follows from Theorem~\ref{thm:1} via a few steps of algebra.\footnote{Proof of the first statement: By \eqref{prop1Abound1}, $|\tr(\rho_A E_z)-\tr(\rho_B E_z)|=|\langle \Psi_A|E_z|\Psi_A\rangle-\tr(\rho_0 E_z)|\leq \|E_z\|/\sqrt{\varepsilon d_0} \, \EC{\leq} \, 1/\sqrt{\varepsilon d_0} < 1/2$. Second statement: $\Psi_A$ and $\Psi_B$ are chosen like independent random variables, so for $(1-\varepsilon)^2$-most pairs $(\Psi_A,\Psi_B)$, \eqref{prop1Abound1} for $B=E_z$ holds for both $\Psi$s, and thus $|\tr(\rho_A E_z) -\tr(\rho_B E_z)|=|\langle \Psi_A|E_z|\Psi_A\rangle - \langle \Psi_B|E_z|\Psi_B\rangle| \leq |\langle\Psi_A|E_z|\Psi_A\rangle -\tr(\rho_0 E_z)| + |\langle \Psi_B|E_z|\Psi_B\rangle -\tr(\rho_0 E_z)| \leq 2/\sqrt{\varepsilon d_0} < 1/2$.} 
Note that it trivially follows that $\rho_A,\rho_B$ cannot be $(1-\delta)$-reliably distinguished for any $\delta$ between 0 and 1/2.
For example, for dimension $d_0=10^{100}$ and any $0<\delta<1/2$, $E_z$ can $(1-\delta)$-reliably distinguish at most the fraction $3.2 \times 10^{-99}$ of the pairs of wave functions.

Corollary~\ref{cor:reliably} should be contrasted with the statement that for large $d_0$ and purely randomly and independently chosen $\Psi_A,\Psi_B\in\SSS(\Hilbert_0)$, the two vectors will be nearly orthogonal, so there exists a POVM that distinguishes them reliably. Note, however, that if we keep this POVM $E$ fixed and consider new, random $\Psi_A,\Psi_B$, then the two yield nearly the same Born distribution for $E$. So, the two statements differ in the order of quantifiers (``for most,'' ``there is''): for most $\Psi_A,\Psi_B$, there is $E$ that reliably distinguishes $\Psi_A,\Psi_B$; but according to Corollary~\ref{cor:reliably}, it is \emph{not} the case that there is $E$ such that for most $\Psi_A,\Psi_B$, $E$ reliably distinguishes $\Psi_A,\Psi_B$. 

Likewise, for every given $\Psi_A\in\SSS(\Hilbert_0)$, there is a POVM $E$ that distinguishes $|\Psi_A\rangle\langle\Psi_A|$ $(1-d_0^{-1})$-reliably from $\rho_0$: just take $E_A=|\Psi_A\rangle\langle\Psi_A|$ and $E_B=I-E_A$.\footnote{Proof: $\tr(\rho_A E_A)=1$ and $\tr(\rho_0 E_B)=1-\tr(\rho_0 E_A)= 1-\langle\Psi_A|\rho_0|\Psi_A\rangle=1-d_0^{-1}$.}

Corollary~\ref{cor:reliably} should also be contrasted with the problem of distinguishing between the case of a mixed state $\rho_0$ and the case of a random pure state $\Psi$ with distribution $u_0$; see \S\ref{sec:otherkinds} for details.

\subsubsection{Not Even Unreliably Distinguishable}
\label{sec:unreliably}

Next, we show that typical vectors in $\Hilbert_0$ are \emph{not even unreliably} distinguishable if the observation isn't too unlikely.

Consider an experiment with POVM $\{E_z:z\in\sZ\}$ and outcome $Z\in\sZ$, and consider two density matrices $\rho_A,\rho_B$. Suppose that for a particular $z$, 
\be
\PPP_A(Z=z)= \tr(\rho_A E_z) = \frac{3}{4}~~~\text{and}~~~\PPP_B(Z=z)= \tr(\rho_B E_z)=\frac{1}{4}.
\ee
Suppose we observed the outcome $z$; then, although it does not distinguish reliably between $\rho_A$ and $\rho_B$ at a reliability level close to 1, one can argue that this outcome at least gives us some probabilistic information, as it favors option $A$ over option $B$ by a factor of 3. We now investigate how strongly our observations can favor $\Psi_A$.

\begin{dfn}
    For any density matrices $\rho_A,\rho_B$ on $\Hilbert$ and any POVM $\{E_z:z\in\sZ\}$, we define the \emph{favoring ratio for $A$ given $z$} to be
    \be
    r_{A|z}:=\frac{\tr(\rho_A E_z)}{\tr(\rho_B E_z)} \,.
    \ee
\end{dfn}

The next statement expresses that in our case, the favoring ratio typically deviates from 1 by merely a negligible amount.

\begin{cor}\label{cor:unreliably}
    Let $\{E_z:z\in\sZ\}$ be any POVM, $z$ any element of $\sZ$, and $\varepsilon>0$. If $\tr(\rho_0 E_z)>\varepsilon$, then for $(1-\varepsilon)$-most $\Psi_A\in\SSS(\Hilbert_0)$, the favoring ratio of $\rho_A=|\Psi_A\rangle\langle\Psi_A|$ over $\rho_B=\rho_0$ obeys
    \be
    1-\frac{1}{\sqrt{\varepsilon^3 d_0}} \leq r_{A|z} \leq 1+ \frac{1}{\sqrt{\varepsilon^3 d_0}} \,.
    \ee
    Furthermore, if also $d_0>9\varepsilon^{-3}$, then for $(1-\varepsilon)^2$-most choices of $\Psi_A, \Psi_B$ in $\SSS(\Hilbert_0)$, the favoring ratio of $\rho_A=|\Psi_A\rangle\langle\Psi_A|$ over $\rho_B=|\Psi_B\rangle\langle\Psi_B|$ obeys
    \be
    1- \frac{2}{\sqrt{\varepsilon^3 d_0}} \leq r_{A|z}\leq 1+ \frac{3}{\sqrt{\varepsilon^3 d_0}} \,.
    \ee
    In both cases, the favoring ratio is close to 1 whenever $d_0\gg \varepsilon^{-3}$.
\end{cor}

This follows from Theorem~\ref{thm:1} through some steps of algebra.\footnote{Proof: 
By hypothesis, $\eta := (\varepsilon d_0)^{-1/2}/\tr(\rho_0 E_z) \leq 1/\sqrt{\varepsilon^3 d_0}$. By Theorem~\ref{thm:1}, $(1-\varepsilon)$-most $\Psi_A$ obey \eqref{prop1Abound1} with $B=E_z$; dividing by $\tr(\rho_0 E_z)$, we obtain that $\langle\Psi_A|E_z|\Psi_A\rangle/\tr(\rho_0 E_z) \in [1-\eta,1+\eta]$. Second statement: By hypothesis, $\eta<1/3$. By Theorem~\ref{thm:1}, for $(1-\varepsilon)^2$-most pairs, both $\Psi$s obey \eqref{prop1Abound1} with $B=E_z$; we obtain the same relation as stated above for $\Psi_A$ also for $\Psi_B$. If $\langle\Psi_A|E_z|\Psi_A\rangle \geq\langle\Psi_B|E_z|\Psi_B\rangle$, then $1\leq \langle\Psi_A|E_z|\Psi_A\rangle /\langle\Psi_B|E_z|\Psi_B\rangle \leq (1+\eta)/(1-\eta)= 1+2\eta/(1-\eta) <1+3\eta$ using $\eta<1/3$. If $\langle\Psi_A|E_z|\Psi_A\rangle \leq\langle\Psi_B|E_z|\Psi_B\rangle$, then $1\geq \langle\Psi_A|E_z|\Psi_A\rangle /\langle\Psi_B|E_z|\Psi_B\rangle \geq (1-\eta)/(1+\eta)=1-2\eta/(1+\eta) \geq 1-2\eta$.}
In words, it means that the outcome $z$, if it is not very improbable, leads to no significant probabilistic preference of $\Psi_A$ over $\rho_0$, or of $\rho_0$ over $\Psi_A$, or of $\Psi_A$ over $\Psi_B$, or of $\Psi_B$ over $\Psi_A$.

What about outcomes 
that \emph{are} very improbable? They should not occur if (as we assume) the theory is correct, unless there are so many possible outcomes that the probabilities of the very improbable ones can add up to a significant value. That situation occurs in the example of 1000 coin tosses mentioned already near \eqref{exepsilon1}, so, unlike the situation considered by Borel, we need to also consider here events \emph{not specified in advance}. That is, we need to reconsider the relevant size of $\varepsilon$ because $\varepsilon$ now appears in two roles: the maximal measure of the exceptional $\Psi$'s and the minimal probability of the improbable observations. So, how small could the probability of an observed outcome be? Like the coin toss example, an experiment can have a large number $L$ of possible outcomes, and if all have equal or similar probability, then these probabilities will be tiny of order $1/L$. If we need to distinguish between $L$ different observations, we can perhaps reach $L=10^{10^{10}}$ by collecting $10^9$ data points, each given by a number with 10 decimal digits. On the other hand, as of this writing (2025), the total storage capacity of all computers in the world is estimated at about $10^{26}$ bytes, so we would not be able to even record the outcome if $L$ exceeded $10^{10^{30}}$, which motivates us to consider,
\be\label{exepsilon}
\text{for example, }\varepsilon=10^{-10^{31}}\,.
\ee
Then, since the number of outcomes would realistically be less than $10^{10^{30}}$, the probabilities less than $\varepsilon$ would add up to less than $10^{-9\cdot 10^{30}}$. Thus, outcomes with probability less than $\varepsilon$ will not be observed given the computational limitations, and still our examples satisfy $d_0 = 10^{10^{80}} \gg 10^{3\cdot 10^{31}} =\varepsilon^{-3}$.

\subsubsection{A Bayesian Analysis}
\label{sec:Bayesian}

Theorem~\ref{thm:1} also allows us to perform a Bayesian analysis, i.e., to mathematically express the probabilistic information provided by the observation of the outcome $z$.
Here, we provide two quantitative bounds on this information. They have the consequence that, if the observation is not too improbable, it cannot yield substantial probabilistic information about the universal quantum state. 
This result invites further analysis of its implications for Bayesian epistemology and convergence to the truth (see \cite{Lin24} for a survey), which we leave for future work.

Suppose the prior probability distribution over quantum states in $\Hilbert_\PH$, $\PPP(\Psi_{t_0})$,  is given by the Past Hypothesis, i.e., by $u_{\PH}$. 
We can show that, if the observation is not too improbable, the posterior probability distribution is very close to the prior. 

\begin{cor}\label{cor:f}
Suppose $d_{\PH}>1/\varepsilon^5$ and $\tr(\rho_{\PH}\,E_z)>\varepsilon$. Consider a prior distribution given by $u_{\PH}$, and let $f(\psi)$ be the density relative to $u_\PH$ of the posterior distribution obtained by Bayesian updating, given that $Z=z$. Then
\be\label{fbound}
1-\varepsilon < f(\psi) < 1+\varepsilon 
\ee
for $(1-\varepsilon)$-most $\psi\in\SSS(\Hilbert_\PH)$. 
\end{cor}

This follows from Theorem~\ref{thm:1} via a short calculation.\footnote{Proof: From $\PPP(\Psi\in d\psi)=u_{\PH}(d\psi)$ for every infinitesimal set $d\psi\subset \SSS(\Hilbert_{\PH})$ and
$\PPP(Z=z|\Psi) = \langle \Psi|E_z|\Psi\rangle$, we find that
$\PPP(\Psi \in d\psi, Z=z) = u_{\PH}(d\psi) \langle\psi|E_z|\psi\rangle$ and
$\PPP(Z=z) = \int_{\SSS(\Hilbert_{\PH})}  u_{\PH}(d\phi) \langle\phi|E_z|\phi\rangle = \tr(\rho_{\PH} E_z)$. Therefore,
$u_{\PH}(d\psi) \, f(\psi) = \PPP(\Psi \in d\psi | Z=z) = u_{\PH}(d\psi) \langle\psi|E_z|\psi\rangle/\tr(\rho_{\PH} E_z)$. By Theorem~\ref{thm:1} for $(1-\varepsilon)$-most $\psi$, $\bigl|\langle\psi|E_z|\psi\rangle -\tr(\rho_{\PH} E_z)\bigr| <\varepsilon^2$, and thus
$\bigl|\langle\psi|E_z|\psi\rangle/\tr(\rho_{\PH} E_z) - 1 \bigr| < \varepsilon^2/ \tr(\rho_{\PH} E_z) < \varepsilon$.} We can also reformulate the result in terms of the probability of any set $S\subseteq \SSS(\Hilbert_{\PH})$ instead of the density function $f$:

\begin{cor}\label{cor:S}
Suppose $d_{\PH}>1/\varepsilon^5$ and $\tr(\rho_{\PH}\,E_z)>\varepsilon$. For any subset $S\subseteq\sphere(\Hilbert_{\PH})$:
\begin{equation}\label{BayesianBound}
    \PPP(\Psi_{t_0}\in S\,|\,Z=z) \in \Bigl[\PPP(\Psi_{t_0}\in S)-2\varepsilon, \PPP(\Psi_{t_0}\in S)+3\varepsilon\Bigr]
\end{equation}
\end{cor}

This follows from Corollary~\ref{cor:f} via standard arguments of probability and integration theory.\footnote{Proof: Let $M\subseteq \SSS(\Hilbert_{\PH})$ be the set where \eqref{fbound} holds. By Corollary~\ref{cor:f}, $u_{\PH}(M)\geq 1-\varepsilon$. Thus, writing $M^c=\SSS(\Hilbert_{\PH}) \setminus M$ for the complement of $M$ and using \eqref{fbound}, $\PPP(\Psi \in S| Z=z)= \int_S u_{\PH}(d\psi) \, f(\psi)\geq \int_{S\cap M} u_{\PH}(d\psi) \, f(\psi) \geq \int_{S\cap M} u_{\PH}(d\psi) \, (1-\varepsilon) = (1-\varepsilon) \, u_{\PH}(S\cap M) \geq (1-\varepsilon) \, (u_{\PH}(S)-u_{\PH}(M^c)) \geq (1-\varepsilon) \, (u_{\PH}(S)-\varepsilon) \geq u_{\PH}(S) -2\varepsilon$. On the other hand, since $f$ is normalized, $\int_{\SSS(\Hilbert_{\PH})} u_{\PH}(d\psi)\, f(\psi)=1$, we have that $\int_{M^c} u_{\PH}(d\psi) \, f(\psi)=1-\int_M u_{\PH}(d\psi) \, f(\psi)\leq 1-\int_M u_{\PH}(d\psi) \,(1-\varepsilon) = 1-(1-\varepsilon) \, u_{\PH}(M) \leq 1-(1-\varepsilon)^2< 2\varepsilon$. Thus, using \eqref{fbound} again, $\PPP(\Psi \in S| Z=z)= \int_S u_{\PH}(d\psi) \, f(\psi) = \int_{S\cap M} u_{\PH}(d\psi) \, f(\psi) + \int_{S\cap M^c}u_{\PH}(d\psi) \, f(\psi) \leq \int_{S\cap M} u_{\PH}(d\psi) \,(1+\varepsilon) + \int_{M^c} u_{\PH}(d\psi) \, f(\psi) < (1+\varepsilon) \, u_{\PH}(S\cap M) + 2\varepsilon \leq (1+\varepsilon) u_{\PH}(S) + 2\varepsilon \leq u_{\PH}(S) + 3\varepsilon$.}

The assumption $d_{\PH}>\varepsilon^{-5}$ is mild; for example for \eqref{exepsilon}, $\varepsilon^{-5}=10^{5\cdot10^{31}}$ is far less than \eqref{exd0}.

The physical consequence of these corollaries will be that, in general, observations reveal very little about the initial quantum state. 

\begin{ex}
    To illustrate that the assumption of large $d_0$ cannot be dropped, we compute $f(\psi)$ explicitly in a case with $d_0=2$: Let $\{|b_1\rangle,|b_2\rangle\}$ be an orthonormal basis of $\Hilbert_0$, and let $E_z=|b_1\rangle \langle b_1|$. Then\footnote{Proof: $\PPP(Z=z|\Psi=\psi)=\bigl| \langle\psi|b_1\rangle \bigr|^2$ by Born's rule, so $\PPP(Z=z,\Psi \in d\psi)=\bigl| \langle\psi|b_1\rangle \bigr|^2 u_0(d\psi)$ and $\PPP(Z=z)=1/2$ by symmetry, so $\PPP(\Psi\in d\psi|Z=z)= 2\bigl| \langle\psi|b_1\rangle \bigr|^2 u_0(d\psi)=f(\psi) \, u_0(d\psi)$, which yields \eqref{f2d}.}
    \be\label{f2d}
    f(\psi) = 2\bigl| \langle\psi|b_1\rangle \bigr|^2 \,,
    \ee
    which varies between 0 and 2 and deviates significantly from 1 for a substantial fraction of $\psi$'s: indeed, it follows from Example~\ref{ex:2} that $f(\psi)$ exceeds 1.8 for 10\%\ of all $\psi$'s.
\end{ex}

\begin{ex}
    To illustrate that the assumption $\tr(\rho_0 E_z)>\varepsilon$ cannot be dropped, we consider an orthonormal basis $\{|b_1\rangle, \ldots, |b_{d_0}\rangle\}$ of $\Hilbert_0$ and the POVM $E_z = |b_z\rangle \langle b_z|$ for $z=1,\ldots,d_0$. Then $\tr(\rho_0 E_z) = 1/d_0$ for any $z$, which is much less than $\varepsilon$ if $d_0>\varepsilon^{-5}$. By the same reasoning as for \eqref{f2d}, $f(\psi) = d_0 |\langle \psi|b_z\rangle|^2$,
   which is far from being constantly 1.\footnote{Proof: It is known that for random $\Psi$ with uniform distribution and large $d_0$, $\langle \Psi|b_z\rangle$ has an approximately Gaussian distribution in the complex plane with mean 0 and variance $\EEE[|\langle \Psi|b_z\rangle|^2]=1/d_0$. Since this distribution gives measure 1/4 to the disk around the origin with radius $\sqrt{\ln(4/3)/d_0}$ and 1/4 to the complement of the disk around the origin with radius $\sqrt{\ln(4)/d_0}$, we have that $f(\psi)<\ln(4/3)<0.288$ for 1/4 of all $\psi$'s and $f(\psi)>\ln(4)>1.38$ for 1/4 of all $\psi$'s.}
\end{ex}

\subsection{Physical Consequences: First Argument for OI}
\label{sec:phys}

We now explain how these mathematical facts imply OI in the sense of observation typicality in OQM, BM, GRW theory, and EQM. We have two arguments, the first of which will be presented now, the second in \S\ref{sec:phys2}; a comparison of the two will be made in \S\ref{sec:framing}.

The first argument is based on the Main Theorem about POVMs (see \S\ref{sec:keythm}), which is known to be valid in OQM, BM, GRW theory, and EQM. The argument puts
together the steps we have laid out already. By the Past Hypothesis, $\Psi_{t_0}$ is typical in $\Hilbert_0$, which has dimension $d_0\geq 10^{10^{80}}\gg \varepsilon^{-5}$ even for $\varepsilon=10^{-10^{31}}$. Suppose we carry out any experiment $\mathscr{E}$; $\mathscr{E}$ is associated, according to the Main Theorem about POVMs, with some POVM $E$; suppose we obtained the outcome $Z=z$ associated with the operator $E_z$. Then according to Corollary~\ref{cor:reliably}, typical quantum states in $\Hilbert_{\macro}$ are not reliably distinguished by this outcome from $\rho_{\macro}$. Suppose further $\tr(\rho_0E_z)>\varepsilon$. By Corollary~\ref{cor:unreliably}, the outcome does not significantly favor one $\Psi$ over another or over $\rho_0$. And by Corollaries \ref{cor:f} and \ref{cor:S}, the Bayesian posterior distribution of $\Psi$ does not significantly differ from the prior. Thus, regardless of which experiment observers are making, the outcome $z$ provides no significant information about $\Psi$. This completes the argument.

This reasoning uses only one outcome $z$ and the associated operator $E_z$; it does not matter how many other $z'$ there are and what the $E_{z'}$ are. OI of typical quantum states is true for any experiment, so it is true for the actual experiment that we happen to perform at an arbitrary time with $z$ standing in for the actual outcome.

In quantum theory with unitary evolution, OI extends to later states too: Since typical $\Psi_t$ from the appropriately evolved subspace $U_t\Hilbert_{\macro}$ lead to nearly identical probability for outcome $Z=z$, the observation of this outcome does not distinguish among different $\Psi_t$.

\subsection{Physical Consequences: Second Argument for OI}
\label{sec:phys2}

Our second argument also works in all of OQM, BM, GRW theory and EQM. It is more direct than the first argument, particularly in BM and the flash version of GRW. We will formulate it first in BM, and afterwards comment on its translation to the other versions of QM.

The argument concerns the configuration $Q_t$ of the universe at an arbitrary time $t$; $Q_t$ is typical with respect to the $|\Psi_t|^2$ distribution, and this distribution is given in terms of the initial wave function $\Psi_{t_0}$ of the universe by a POVM (in fact, a PVM (projection-valued measure)) $P$,
\be
 \int_\Omega dq \; |\Psi_t(q)|^2 = \langle\Psi_{t_0}|P(\Omega)|\Psi_{t_0}\rangle
\ee
for any subset $\Omega$ of the configuration space of the universe. This POVM $P$ is the Heisenberg-evolved configuration observable, i.e., the projection-valued measure jointly diagonalizing all position operators at time $t$, or $P(\Omega)=U_t^\dagger 1_\Omega U_t$, where $1_\Omega$ is the multiplication operator multiplying by the characteristic function of $\Omega$.

Now $Q_t$ comprises the exact positions of all particles in the universe, so observers inside a Bohmian universe cannot possibly measure $Q_t$; at best, they can measure a very coarse-grained version of some components of $Q_t$. But the crucial point here is that any information available to the inhabitants of a Bohmian universe at time $t$ will be recorded in $Q_t$, in fact in whether $Q_t$ lies in some (suitably coarse, macroscopic) set $\Omega$. For example, observations (and conclusions drawn from them) will be reported in books and journals, and thus in the configuration of ink particles. (For more detailed discussions, see \citep[pp.~83-85]{Alb15}.) More precisely, we make use of the following
\begin{description}
    \item[Necessary Condition]
    If an event is \emph{known to inhabitants of a Bohmian universe at time $t$} then it is of the form $Q_t \in \Omega$ for some set $\Omega$.
\end{description}

The argument for OI in BM consists of applying Corollaries~\ref{cor:reliably}--\ref{cor:S} to the POVM $\{E_A,E_B\}=\{P(\Omega),P(\Omega^c)\}$ for some subset $\Omega$ of the configuration space $\RRR^{3N}$ of the universe and its complement $\Omega^c=\RRR^{3N}\setminus \Omega$. As a consequence of the necessary condition just mentioned and Corollaries~\ref{cor:reliably}--\ref{cor:S}, no event known to the inhabitants can reliably distinguish $\Psi_A$ from $\rho_0$, or can significantly favor $\Psi_A$ over $\rho_0$ or $\Psi_B$, or can substantially alter the Bayesian posterior distribution of $\Psi_{t_0}$ (for the last two disjuncts provided $\tr(\rho_0 P(\Omega))>\varepsilon$, i.e.,  the event $Q_t\in\Omega$ is not too improbable on average). Thus, at no time $t$ can inhabitants of a Bohmian universe have significantly more information about $\Psi_{t_0}$ than what is already provided by the Past Hypothesis. This completes the argument.

This conclusion covers also the possibility that observers might make observations (say, with telescopes) or experiments, as the outcomes of these observations or experiments will be recorded in the configuration of some particles (e.g., when scientists publish their findings). It also covers the possibility of making observations or experiments at several times $t_1,t_2,\ldots,t_k$, since the outcomes will be recorded and thus can be read off from $Q_t$ at $t\geq t_k$. 

We remark that the precise microscopic trajectories of all Bohmian particles may contain more information about $\Psi_{t_0}$, but it is known 
that observers inside a Bohmian universe cannot measure positions at several times without introducing decoherence and thereby changing the trajectory. Therefore, they do not have access to the full trajectories, which makes understandable why the dependence of the trajectories on $\Psi_{t_0}$ is compatible with OI as just derived.

Let us summarize the second argument in a different way. A key fact in BM that leads to OI is this:
\be\label{OIBM}
\begin{minipage}{0.7 \textwidth}
    \textit{for any subset $\Omega$ of configuration space, the probability of the event $Q_t\in\Omega$, i.e., $\int_\Omega dq \; |\Psi_t(q)|^2$, is nearly independent of $\Psi_{t_0}$ for typical $\Psi_{t_0}\in\SSS(\Hilbert_{\macro})$.}
\end{minipage}
\ee
Corollaries~\ref{cor:reliably}--\ref{cor:S} just provide quantitative statements (in the more general framework of POVMs) of how this near-independence makes $\Psi$ observationally indistinguishable from $\rho_0$ or other $\Psi$s. They show that 
that even complete information about $Q_t$ would not help with distinguishing between typical $\Psi_{t_0}$ and $\rho_0$ (or between two typical wave functions); \textit{a fortiori}, limited information about $Q_t$ does not help.

In \textbf{OQM} \EC{(that is, textbook QM; see \citealp[sec.~4]{Barrett2019} for a representative summary)}, the second argument can be applied as well (although it is a bit less clear because the ontology of OQM is less clear): after all, the Heisenberg cut (between system and apparatus) can always be moved so as to enlarge the system, and the $|\Psi_t|^2$ distribution in configuration space will also in OQM determine the distribution of macroscopic records (say, in books or journals) at time $t$. Thus, the macroscopic records at $t$ provide almost no information about $\Psi_{t_0}$.

Concerning \textbf{EQM}, we have reservations about the justification of the use of probabilities  (\citealp[sec.~6]{BarrettSEP2023}; \citealp[sec.~3.5.5]{tumulka2022foundations}); but to the extent that they can be justified, our second argument for OI applies as well.

In \textbf{GRW theory}, the second argument can be applied in a slightly modified variant. To this end, we
consider GRWf, the GRW theory with a flash ontology in spacetime (\citealp[sec.~11]{Ghi07}; \citealp[sec.~3.3.4]{tumulka2022foundations}). Let $F_t$ denote the pattern of flashes up to $t$, with $t>t_0$. A key fact about OI in GRWf is this:
\be\label{OIGRWf}
\begin{minipage}{0.7 \textwidth}
    \textit{for any subset $M \in \cup_{n=0}^\infty\RRR^{4n}$, the probability of the event that  $F_t \in M$ is nearly independent of $\Psi_{t_0}$ for typical $\Psi_{t_0}\in\SSS(\Hilbert_{\macro})$.}
\end{minipage}
\ee
This follows from Theorem~\ref{thm:1} by taking the POVM $E$ to be the POVM governing the joint distribution of all flashes \citep[sec.~5.1.1 and 7.8]{tumulka2022foundations}. As for the configuration in BM, observers in a GRWf universe do not have access to the full pattern of flashes, $F$, at most to a coarse-grained version of it. But again, we use the fact that any record would have to be encoded in the flashes (as are the positions of pointers or the shape of printed letters). More precisely, we use the following
\begin{description}
    \item[Necessary Condition]
    If an event is \emph{known to inhabitants of a GRWf universe at time $t$} then it is of the form $F_t \in M$ for some set $M$.
\end{description}
Now Corollaries~\ref{cor:reliably}--\ref{cor:S} show that no 
such event can reliably distinguish $\Psi_A$ from $\rho_0$, or can significantly favor $\Psi_A$ over $\rho_0$ or $\Psi_B$, or can substantially alter the Bayesian posterior distribution of $\Psi_{t_0}$, provided $\tr(\rho_0 E(M))>\varepsilon$. Thus, at no time $t$ can observers in a GRWf universe have significant information about $\Psi_{t_0}$. This completes the argument.

A difference to the Bohmian case is that the exact \emph{history} of the universal configuration in BM may provide sufficient information to determine $\Psi_{t_0}$, whereas in GRWf even the exact history of all flashes is of no help. Thus, GRWf provides, in a sense, a stronger kind of limitation to knowledge than BM.

Another difference to BM is that inhabitants of a GRWf universe living at time $t$ may very well find out a lot about the wave function $\Psi_t$ at their time (in contrast to the initial wave function $\Psi_{t_0}$). After all, $\Psi_t$ is not unitarily evolved from $\Psi_{t_0}$ but has collapsed; for example, if we find Schr\"odinger's cat alive then its wave function will have collapsed to (approximately) the wave function of a live cat. (Note that this still does not give us new information about $\Psi_{t_0}$.)

\subsection{Remarks}
\label{sec:remarks}

\subsubsection{Experiment Framing vs $|\Psi_t|^2$ Framing}\label{sec:framing}

Our first remark concerns a comparison of the two arguments given in \S\ref{sec:phys} and \S\ref{sec:phys2}. Although both used the Corollaries \ref{cor:reliably}--\ref{cor:S}, they used them in different ways: the first was framed in terms of experiments, the second in terms of the $|\Psi_t|^2$ distribution in configuration space. The first argument pretends that the observers are somehow not part of the universe. Although we do not think that this approach leads to any wrong conclusions here, we note that the second argument is clear-cut and the first less so, as various questions can arise about the first: Is it not usually assumed that observers and their apparatus are not entangled with the object they are observing? And is that not violated in the relevant observations of the universe? Given that in OQM, observations can collapse wave functions \EC{(\citealp[pp.42-43]{Barrett2019})}, should not $E_z$ be applied to a collapsed wave function instead of $\Psi_{t_0}$? While none of these objections to the first argument withstands scrutiny, the second argument makes it much more transparent that the conclusion, OI, remains valid. The second argument also makes it clear that repeating the experiment does not allow the observer to measure the probability more precisely; after all, records of all experiments are included in the configuration at time $t$, and the whole universe cannot be repeated, of course.

\subsubsection{Other Kinds of Indistinguishability}
\label{sec:otherkinds}

As already remarked, the question we ask is different from the question, given two states $\Psi_A,\Psi_B$, is there a POVM that will distinguish them? It is also different from the question, can we empirically distinguish the possibility that the universe is in the mixed state $\rho_0$ from the possibility that it is in a random pure state $\Psi$ with distribution $u_0$? The answer to the latter question is trivially ``no'' for any value of $d_0$ (large or not), as any observation $z$ with POVM element $E_z$ has probability $\tr(\rho_0 E_z)$ in the mixed case and probability $\int_{\SSS(\Hilbert_0)} u_0(d\psi) \langle\psi|E_z|\psi\rangle$ in the random pure case, and the two probabilities are always exactly equal.

Our results convey that the Born distribution $\langle \Psi|E_z|\Psi\rangle$ is close to $\tr(\rho_0 E_z)$ \emph{for most $\Psi$}, whereas the impossibility to distinguish the mixed case from the random pure case arises from the fact that $\langle \Psi|E_z|\Psi\rangle$, when \emph{averaged over $\Psi$}, yields $\tr(\rho_0 E_z)$. The difference between our results and the impossibility to distinguish the mixed case from the random pure case is perhaps most easily visible in the Bayesian statement of \S\ref{sec:Bayesian}.

\section{Comparisons with Known Results}
\label{sec:known}

There are several results about limitation to knowledge about the quantum state.  For a recent survey, see \cite[ch.5]{tumulka2022foundations}. Here we highlight a few: 
(1) Because of the Main Theorem about POVMs, we cannot measure the wave function of a given system with useful precision. 
    Relatedly, because of the no-cloning theorem, we cannot reliably copy the wave function of a given system. 
    (2) It can be shown \citep{Tum22a} that in any ontological theory of quantum mechanics, it is impossible to measure the ontic state. Thus, whatever the true theory of quantum mechanics may be, there must be facts that cannot be empirically determined.
    (3) Because of the Pusey-Barrett-Rudolph (PBR) theorem \citep{pusey2012reality}, two different ensembles of wave functions with the same density matrix are physically distinct. However, they are observationally indistinguishable by all possible observations directly on the ensembles.

The last item also asserts a kind of OI, but differs from our result here in that it provides a condition (equal density matrix) under which two situations (ensembles of wave functions) are indistinguishable, whereas we show here that most wave functions from $\Hilbert_\PH$ are indistinguishable.
All three items are compatible with substantive learning about the quantum state: \EC{one might still expect to rule out many possibilities or sharply update one's credences. Observation typicality shows this hope is misplaced, as observations eliminate only a tiny fraction of candidates.}

In the philosophy of general relativity (GR), \cite{manchak2009can, manchak2011physically} showed that for every spacetime model in a certain class, there exists another that is physically distinct (different global structure) yet (locally) observationally indistinguishable.\footnote{For discussion, see \cite{Beisbart2009, Norton2011, Butterfield2012, CintiFano2021}.} \EC{Thus both GR and QM imply OI, though in different ways: Manchak proves the existence of OI counterparts for each model of GR by a ``cut-and-paste'' construction (sometimes criticized as physically unreasonable, cf. \cite{CintiFano2021}), whereas we use a probabilistic method to show that typical models of QM in high-dimensional Hilbert space are OI from each other. Unlike Manchak’s result, ours is not compatible with observationally ruling out many possibilities. Moreover, our results apply to \emph{any} observation, local or otherwise. }

\section{Philosophical Implications}
\label{sec:implications}

We sketch four potential philosophical implications of observation typicality. Our aim here is exploratory, not prescriptive.

First, observation typicality imposes severe limitations on knowledge in a quantum universe. They are in-principle limitations that cannot be overcome, regardless of technological advancements, except in cases where the relevant provisos do not apply.

Second, the flip side is that observation typicality allows us to know a great deal about the observable properties of the universal quantum state. For any given observation, nearly all universal quantum states are indistinguishable from the normalized projection $\rho_{\PH}$. 
Hence, fixing the Past Hypothesis almost completely determines the probabilistic predictions of typical quantum states in $\Hilbert_0$.

Third, observation typicality may conflict with positivism, the idea that a statement is unscientific or meaningless if it cannot be tested experimentally, and a variable is not well-defined if it cannot be measured. Given the PBR theorem \citep{pusey2012reality}, and the various solutions to the quantum measurement problem, we have good reasons to accept that the universal quantum state $\Psi$ is an objective feature of a quantum universe, yet it is observationally underdetermined because of observation typicality. This is ironic, to echo a point made by \cite{cowan2016epistemology}: positivism is sometimes defended using examples from quantum mechanics,  yet quantum mechanics itself reveals that a central object in a quantum universe---the universal quantum state---is well-defined, objective, but fundamentally unmeasurable. 

Finally, our results impact quantum state realism (see \cite{chen2019realism} for a survey). One view treats the universal quantum state as a physical entity---a field constituting one of the universe's basic building blocks \citep{AlbertEQM, ney2021world}. Yet observation typicality implies limits to even imprecisely measuring it, challenging the empiricist expectation that fundamental constituents of the universe are in principle measurable (at least imprecisely). Alternatively, if we interpret the universal quantum state as a physical law (akin to the classical Hamiltonian; see \cite{goldstein2013reality}), we may naturally expect it to be constrained by observational evidence but remain vastly underdetermined, much like other physical laws.

Quantum state realism leaves open the question of whether it must necessarily be a pure state, represented by a wave function, or if it could be a mixed state, represented by a density matrix. Following \cite{chen2018IPH, chen2024DMR}, we refer to the first option as \textit{wave function realism} (WFR) and the second as \textit{density matrix realism} (DMR). 
Not only are typical individual wave functions OI from each other, but they are also OI from $\rho_0$, which is mixed. This may offer a defeasible reason to prefer DMR over WFR. If typical pure states $\Psi_{t_0}$ compatible with the Past Hypothesis yield nearly identical predictions as $\rho_{\PH}$, why not adopt the Wentaculus version of DMR and postulate the Wentaculus density matrix $\rho_{\PH}$ instead? After all, $\rho_{\PH}$ can be regarded as much simpler than any typical $\Psi_{t_0}$ while maintaining near predictive equivalence.

\section{Conclusion}
\label{sec:conclusion}

We have shown that a central object in \EC{several leading formulations of} quantum theory---the quantum state of the universe---is effectively hidden from observation if it is typical. Typical quantum states in a high-dimensional Hilbert subspace $\Hilbert_{\macro}$ are observationally indistinguishable from the density matrix $\rho_{\macro} = P_{\macro}/d_{\macro}$ and, consequently, from each other. No observation will yield substantial information about $\Psi_{t_0}$---not reliably and (unless the observation is extremely unlikely) not even unreliably. These results amount to the strongest epistemic limits we know for a quantum universe. If the wave function of our own universe is typical in $\Hilbert_0$, our observational data alone will reveal very little about which quantum state it is in. Nature, it seems, is far more secretive than we have realized.

\bigskip

\textbf{Acknowledgment.} We thank Sheldon Goldstein for his extensive feedback and for suggesting the term ``observation typicality.'' For helpful discussions, we also thank Emily Adlam, David Albert, Fabio Anza, Patricio Avila Cardenas, Jeff Barrett, Jeremy Butterfield, Craig Callender, Jennifer Carr, Silvia De Bianchi, Branden Fitelson, Hans Halvorson, Alan H\'ajek, Tyler Hildebrand, Hsin-Yuan Huang, Nick Huggett, Christopher Jarzynski, Andrew Jordan, Joel Lebowitz, Matthew Leifer, Barry Loewer, JB Manchak, Tim Maudlin, Travis McKenna, Kerry McKenzie, Marcos Rigol, Daniel Rubio, Charles Sebens, Josiah Sinclair, Shelly Yiran Shi, Tony Short, Stefan Teufel, Mordecai Waegell, Isaac Wilhelm, Nicole Yunger Halpern, the CalTech Philosophy of Science Group, the SoCal Quantum Foundations Hub, the SoCal Philosophy of Physics Group, the Metro Area Philosophy of Science Group, the COSMOS research team, the UC San Diego Philosophy of Physics Group, the participants at the 2024 Quantum Thermodynamics Workshop at the University of Maryland, College Park, the 2025 APS March Meeting, the 2025 Foundations of Thermodynamics Workshop at NTU, and two anonymous referees at \textit{The British Journal for the Philosophy of Science}. 

\textbf{Funding.} EKC is supported by Grant 63209 from the John Templeton Foundation. The opinions expressed in this publication are those of the authors and do not necessarily reflect the views of the John Templeton Foundation.

\textbf{Conflict of interests.} The authors declare no conflict of interest.



\end{document}